\documentclass[conference]{IEEEtran}
\IEEEoverridecommandlockouts
\usepackage{cite}
\usepackage{amsmath,amssymb,amsfonts}
\usepackage{algorithmic}
\usepackage{graphicx}
\usepackage{textcomp}
\usepackage[table]{xcolor}
\def\BibTeX{{\rm B\kern-.05em{\sc i\kern-.025em b}\kern-.08em
    T\kern-.1667em\lower.7ex\hbox{E}\kern-.125emX}}
    
\usepackage{array}
\usepackage{here}
\usepackage{CJKutf8}
\usepackage{multirow}

\ifCLASSOPTIONcompsoc
\usepackage[caption=false,font=normalsize,labelfon
t=sf,textfont=sf]{subfig}
\else
\usepackage[caption=false,font=footnotesize]{subfi
g}
\fi
\usepackage{url}
    
\begin{document}

\title{Retrospective Analysis of Controversial Subtopics on COVID-19 in Japan
\thanks{This work was supported by JST grants 19K20413/21K17859, Japan.}
}


\author{
\IEEEauthorblockN{
    Kunihiro Miyazaki\textsuperscript{\rm 1}, Takayuki Uchiba\textsuperscript{\rm 2}, Fujio Toriumi\textsuperscript{\rm 1}, Kenji Tanaka\textsuperscript{\rm 1}, Takeshi Sakaki\textsuperscript{\rm 1,3}
}
\IEEEauthorblockA{
    \textsuperscript{\rm 1} \textit{School of Engineering, The University of Tokyo, Tokyo, Japan} \\
    \textsuperscript{\rm 2} \textit{Sugakubunka Co., Ltd., Tokyo, Japan} \\
    \textsuperscript{\rm 3} \textit{R\&D dept., Hotto Link Inc., Tokyo, Japan}
}
\IEEEauthorblockA{
miyazaki@ioe.t.u-tokyo.ac.jp,
takayuki.uchiba@sugakubunka.com,\\
tori@sys.t.u-tokyo.ac.jp,
tanaka@tmi.t.u-tokyo.ac.jp,
sakaki@ipr-ctr.t.u-tokyo.ac.jp
}
}

\maketitle

\begin{abstract}
For efficient political decision-making in an emergency situation, a thorough recognition and understanding of the polarized topics is crucial.
The cost of unmitigated polarization would be extremely high for the society; therefore, it is desirable to identify the polarizing issues before they become serious.
With this in mind, we conducted a retrospective analysis of the polarized subtopics of COVID-19 to obtain insights for future policymaking.
To this end, we first propose a framework to comprehensively search for controversial subtopics.
We then retrospectively analyze subtopics on COVID-19 using the proposed framework, with data obtained via Twitter in Japan.
The results show that the proposed framework can effectively detect controversial subtopics that reflect current reality.
Controversial subtopics tend to be about the government, medical matters, economy, and education; moreover, the controversy score had a low correlation with the traditional indicators--scale and sentiment of the subtopics--which suggests that the controversy score is a potentially important indicator to be obtained.
We also discussed the difference between subtopics that became highly controversial and ones that did not despite their large scale.
\end{abstract}


\section{Introduction}
Emergency situations warrant quick and accurate decisions on a variety of issues.
During the situations such as pandemics or earthquakes, people’s concerns, anxieties, and needs have been shifting constantly.
This has increased the necessity for government decision-makers to communicate effectively with people and take the right measures at the right time; otherwise, people are likely to become confused and the solution to the ensuing problems becomes difficult to obtain.
For example, during pandemics, if the development and provision of vaccines are delayed owing to opposition, it could lead to major obstructions in people's lives and economic operations~\cite{burki2019vaccine}.
To prevent such confusion, the government needs to properly conduct risk communication~\cite{national1989improving}, where the government closely considers people's voices and ensures that they are reflected in the government’s political operations.

Polarization is an aspect requiring the attention of the government when listening to people's voices.
Polarization is a state in which people's opinions are severely divided~\cite{conover2011political,schmidt2018polarization,williams2015network}, and those adhering to one view rarely communicate with those with differing views.
When polarization arises in social matters, it is difficult to address it through communication because people are divided by their reinforced beliefs, and the cost incurred by society because of these problems is extremely high~\cite{makridis2020real}.
Moreover, the polarity tends to accelerate if left unmitigated~\cite{sasahara2021social}; hence, it is desirable to accurately identify polarization features and mitigate them before the problem becomes severe~\cite{nelimarkka2019re}.

With this in mind, we conducted a retrospective analysis of the polarized subtopics of COVID-19.
During the COVID-19 pandemic, there was a perceived polarization on a variety of issues, such as perspectives on vaccine and economic policy.
Understanding these controversial subtopics is crucial for efficient political decision-making.
Therefore, analyzing which subtopics are polarized in an emergency situation such as COVID-19 can provide important insights for future policy decision-making.

The problem, however, is that there is still scanty research to comprehensively grasp polarizing subtopics in the society.
Indeed, social polarization is not a new research topic, and various studies have been conducted on it.
For example, one of the most relevant to our aim is the quantification of polarization~\cite{garimella2018quantifying,guerra2013measure}. 
In this context, polarization is referred to as a controversy or echo chamber effect~\cite{garimella2018quantifying,gillani2018me}.
Effective methods have already been proposed to quantify polarization, and many of these quantify the fragmentation of clusters in a network structure~\cite{garimella2018quantifying,guerra2013measure}.
However, these methods are aimed at quantification for a single topic (and not multiple topics) and not investigating social media at large.
Similar research was conducted by Coletto et al.~\cite{coletto2017motif}, where they searched for local discussions in a network created for the main discussion; however, it is a method that searches for intensive interactions rather than fragmentation.

Therefore, we first propose a framework to exhaustively search for controversial subtopics; specifically, we created a shortlist of subtopics by filtering the number of posts on each subtopic, and then quantifying the controversy for each subtopic to extract the ones with a high controversy score.
For the proposed framework, we decided to leverage an existing method because an effective method for controversy quantification has already been proposed in previous research.
We also note that this framework can be applied to topics other than politics, such as marketing, where risk communication is necessary when it comes to viral marketing of some products~\cite{woerdl2008internet}, which would further enhance the importance of developing a framework for detecting controversial subtopics.

We then retrospectively analyze COVID-19 using the proposed framework, with data obtained via Twitter in Japan.
Setting COVID-19 as the main topic of analysis, we examined the subtopics (e.g., vaccines) that fall within the scope of the polarization trend.
The analysis comprises three parts.
The first part is to verify whether the method for controversy quantification works appropriately for our pre-specified subtopics. 
To verify its suitability, we chose well-known subtopics such as the Olympics or vaccines, and compared the results with the news on the mainstream media.
The research question (RQ) is:
\begin{quote}
    RQ1: Does the method for controversy quantification work properly?
\end{quote}
In the second part, we explored controversial subtopics using the proposed framework without pre-specifying the subtopics, and examined their characteristics.
The research question is:
\begin{quote}
    RQ2: Which COVID-19-related subtopics were controversial?
\end{quote}
We then analyzed these subtopics from the perspectives of scale and sentiment, which are traditional indicators used in social media analysis~\cite{batrinca2015social}, and examined the relationships between the controversy scores and these indicators. The research question is as follows:
\begin{quote}
    RQ3: What was the scale and sentiment of the controversial subtopics?
\end{quote}

Consequently, we found that the proposed framework could well detect controversial subtopics that accurately reflect reality.
In addition, the subtopics that are highly controversial are about the government, medical matters, the economy, and education.
Moreover, the controversy score had a low correlation with the scale and sentiment of the subtopics.
This suggests that, given the political importance of remedying public polarization, the controversy score is an important indicator to understand its extent along with the scale and sentiment.
Besides, in Discussion section, we gave our consideration on the difference between subtopics that became highly controversial and ones that did not despite their large scale.

Our contributions are as follows: 
\begin{itemize}
    \item We proposed a framework to comprehensively detect controversial subtopics on social media, and verified that it could detect controversial subtopics that adequately reflect reality. 
    \item We detected the subtopics associated with COVID-19 that were controversial in Japan and found that they are mainly about the government, medical field, economy, and education.
    \item We compared the controversy scores with traditional indicators--scale and sentiment--and found a small correlation between them, which indicates their potential as independent indicators.
\end{itemize}

\section{Dataset}
We used the COVID-19-related dataset.
We used Twitter's standard API to collect Japanese tweets by setting the language for search as Japanese.
The search query mixed Japanese and English words (for Japanese words, the translated English words are in brackets):
\begin{CJK}{UTF8}{ipxm}
``コロナウイルス/コロナウィルス (coronavirus),'' 
``コロナ (corona),''
``ウイルス/ウィルス (virus),''
``新型肺炎 (new type pneumonia),'' 
``新型 (new type),'' 
``肺炎 (pneumonia),''
``武漢 (Wuhan),''
``covid,''
``covid-19,''
``covid19,''
``corona,''
``virus,''
``coronavirus,''
``wuhan''.
\end{CJK}
The period was from February 1 to August 31, 2020.
The tweets, including retweets (RTs), totaled 245,558,103.

\section{Methodology}\label{Methodology}
\begin{figure}[htbp]
\centerline{\includegraphics[width=0.92\linewidth]{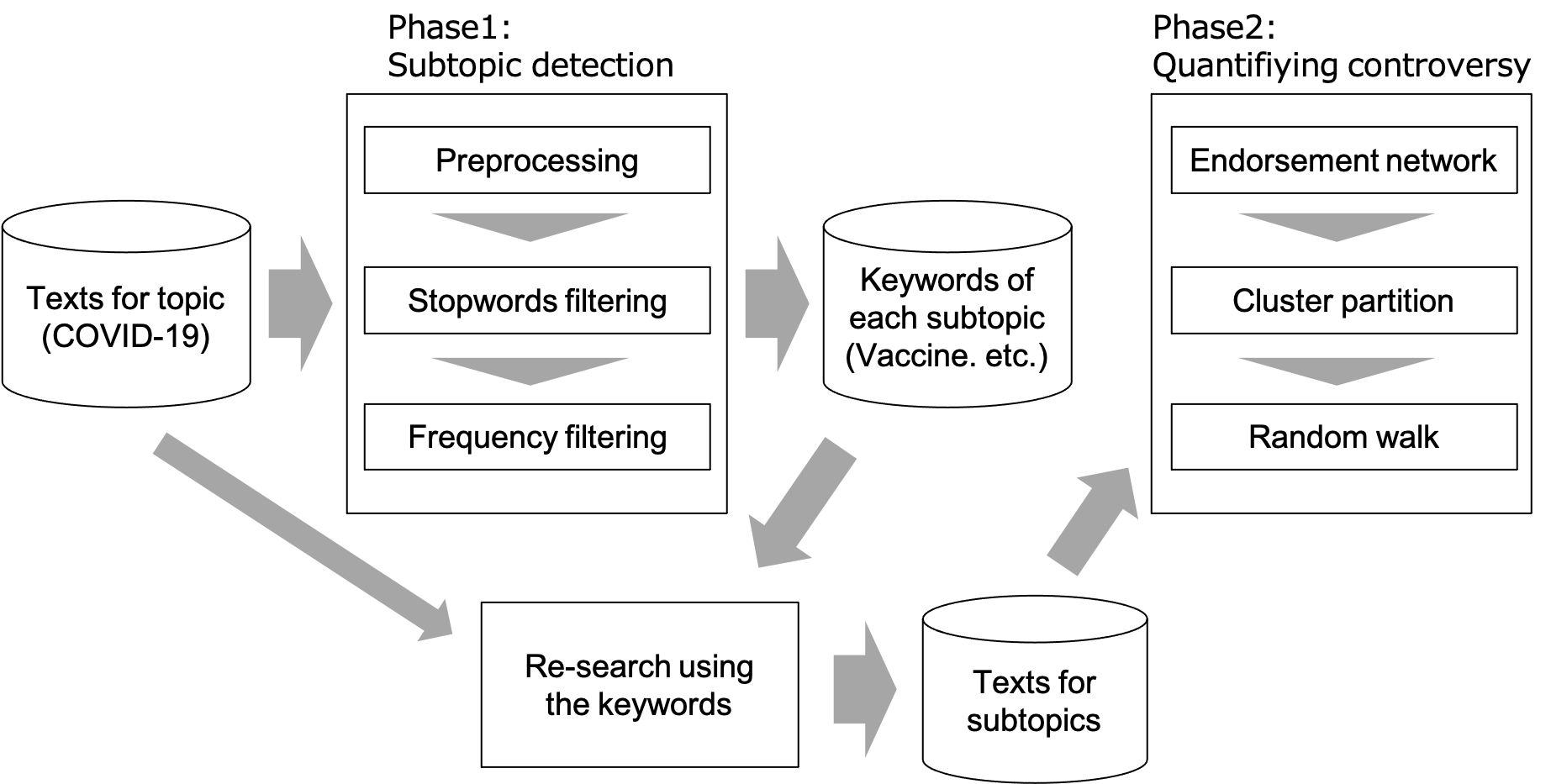}}
\caption{Overview of the proposed framework.}
\label{overview}
\end{figure}

We propose a framework to discover controversial subtopics.
An overview is given in Figure~\ref{overview}.

\noindent This framework consists of two main phases: subtopic detection (phase 1) and controversy quantification (phase 2).
In subtopic detection, we obtain single words to represent the subtopics.
Between phases 1 and 2, we filter the original database using these words as a query and pass the resulting subset of tweets associated with each subtopic to phase 2.
For quantifying controversy, we assign the controversy scores to each subtopic to discover highly controversial subtopics.

\subsection{Phase 1: subtopic dection}
We required two conditions for subtopic detection.
First, for interpretability and search efficiency, one query word sufficiently represents a subtopic.
For example, if the subtopic concerns vaccines, it is desirable to express the subtopic with a single word such as ``vaccine'' excluding ``vaccination'' or ``vax.''
This condition enables us to avoid interpreting the acquired subtopics manually, as is often required when using topic modeling techniques~\cite{murdock2015visualization}--thus, the full automation of the framework is possible, which is helpful in the practical application.
Second, a certain number of query search results is necessary for phase 2.
As reported in previous research~\cite{garimella2018quantifying}, the method for quantifying controversy does not work well in a small network.
Our preliminary experiments found that it is desirable to have more than 800 nodes in the network to obtain robust results.

To fulfill these conditions, we decided to take the following three steps that are indicated in phase 1's box in Figure~\ref{overview}.
First, by preprocessing, we retained only nouns.
In addition, we removed meaningless words (e.g., ``you'') using the ordinary stopword list.
As for the tokenization of Japanese sentences, we used MeCab~\cite{kudo2006mecab}, a Japanese morphological analysis library to obtain nouns from the Japanese dictionary NEologd~\cite{sato2017implementation}.
Second, we created a custom stopword list for further filtering.
Here, we set up the following stopwords to leave the COVID-19's subtopics.
\begin{itemize}
    \item the topic itself (in this case, COVID-19) (``virus'', ``infection'', etc.)
    \item news-related words (``news'', ``breaking'', etc.)
    \item announcements (calls to action such as ``everyone'', etc.)
    \item region, personal name, time, or person (e.g., ``man'')
    \item words that have no meaning (e.g., ``lol'')
\end{itemize}
Third, we sorted the words by frequency and extracted the top-N words to a shortlist for quantifying controversy in phase 2.
For this study, we set the N as 50.

Here, we did not adopt the probabilistic modeling or clustering approaches such as K-means or LDA for subtopic detection despite them being topic detection methods because both can retrieve trivial words, which exacerbates the robustness and interpretability of subtopics. 
Also, these methods are said not to perform well with short sentences such as tweets~\cite{hong2010empirical}.
Actually in our preliminary experiments, we tried these techniques,
but satisfactory results were not obtained.
Thus, as representations of subtopics, we decided to use single words, which are easily interpretable and not associated with noise such as trivial words.
We also decided to use the top frequently used words because of having to guarantee a certain number of tweets for subtopics, the second condition for phase 1.
Then, our task was to retrieve topical words worth picking up from top frequent words.
Here, we found that this task was almost analogous with carefully crafting stopwords as described above.
From the preliminary experiments, as long as we removed trivial words that did not pass to phase 2 (already listed as the crafted stopword list), all the remaining words were independently somewhat topical and worthy of retrieval.

\subsection{Phase 2: quantifying controversy}
We employed the method suggested by Garimella et al.~\cite{garimella2018quantifying} because this method is the state of the art for quantifying controversy and was proved its superiority compared with other methods in~\cite{garimella2018quantifying}.
This method measures controversy by quantifying the connectivity of the two clusters when the network is bisected.
If the connectivity of the two clusters is low, the controversy will be high.
For the connectivity of two clusters, a random walk with restart is employed, and the random walk starts from each cluster to quantify how easy it is to move from one cluster to another.
This random-walk controversy score ($RWC$) is formulated as follows,
\begin{equation}
    RWC = P_{XX^{+}}P_{YY^{+}}-P_{XY^{+}}P_{YX^{+}},
\end{equation}
where $P_{AB^{+}}$ indicates the possibility of moving from the nodes in cluster $A$ to the high-degree nodes in cluster $B^{+}$.
$RWC$ captures the difference in the probability of staying on
the same side and crossing the boundary and takes a value from -1 to 1.
For dividing the network, we employed the METIS algorithm~\cite{karypis1997metis} following the original paper~\cite{garimella2018quantifying}, which splits the network into two clusters of almost equal size.

To obtain $RWC$, we created a network with the following conditions.
The nodes indicate users.
We created an edge between users with more than two RTs (including mutual RTs), which incorporate the meaning of the endorsement into the edges more robustly~\cite{garimella2018quantifying}.
After creating the network, we applied $k$-core decomposition ($k$=2) to exclude users with only weak connections to the primary discussions~\cite{alvarez2006large}. 
Then, we extracted the maximum connected components in the network to remove tiny network fractions.

As mentioned, we set the threshold for the number of nodes to 800.
Additionally, since tweet volumes differ by subtopic, in practice we had to set the appropriate period for obtaining tweets according to the subtopics.
In this study, however, we set a fixed term for comparatively analyzing the extracted subtopics.
We adopted a conservative period of one month.

\section{RQ1: Does the method for controversy quantification work properly?}\label{Validation}
Before using the subtopic detection method, we manually selected subtopics on COVID-19 that were well-known to the public to validate whether the method for quantifying controversy well reflected reality.
\begin{CJK}{UTF8}{ipxm}
We selected the following five words as pre-specified subtopics: ``オリンピック (Olympic),'' ``ワクチン (Vaccine),'' ``GoTo'', ``発熱 (Fever),'' ``死者 (Fatality)''.
\end{CJK}


\begin{CJK}{UTF8}{ipxm}
We assumed ``オリンピック (Olympic),'' ``ワクチン (Vaccines),'' and ``GoTo'' are controversial subtopics.
\end{CJK}
``Olympic'' is a subtopic that concerns holding the sporting event.
In Japan, after the COVID-19 outbreak, there has been a huge discussion over the postponement or cancellation of the biggest global event.
Vaccine concerns the healthcare policy. Recent studies have shown vaccination is one of the most controversial topics online~\cite{schmidt2018polarization} especially in Japan~\cite{wec2020us}. 
``GoTo'' indicates a unique Japanese policy for supporting the travel industry, Go To Travel, which allows travelers to obtain discounts when they travel.
The policy was announced in May 2020 and implemented from July to December.
Although the Japanese government managed to execute this policy, there was criticism that the policy spread the COVID-19 infection.
We also assumed fever and fatalities as non-controversial subtopics because they indicated the factual situation with which opinions have a lower probability of association.
Added to these five words, we quantified controversy without any query as the reference for COVID-19 itself (shown as ``ALL'').

Table~\ref{validation_result} shows the controversy scores for each subtopic by month.
There are cells with dashes where the number of users did not reach the threshold of 800.
We set the threshold of the high/low controversy score to 0.3 based on~\cite{garimella2018quantifying}, and ones that exceeded this score are shown in bold.
\begin{CJK}{UTF8}{ipxm}
\begin{table*}[htbp]
\centering
\caption{Controversy scores for pre-specified subtopics in each month}
\label{validation_result}
\begin{tabular}{cccccccc}
\hline
Subtopic
   & \multicolumn{1}{c}{Feb.}                                             & \multicolumn{1}{c}{Mar.}                                             & \multicolumn{1}{c}{Apr.}                                           & \multicolumn{1}{c}{May.}                                                 & \multicolumn{1}{c}{Jun.}                                              & \multicolumn{1}{c}{Jul.}                                                 & \multicolumn{1}{c}{Aug.}                                                 \\ \hline
オリンピック (Olympic)  & 0.115   & \textbf{0.314}   & 0.110   & -       & -       & \textbf{0.389}   & -              \\
ワクチン (Vaccines)  & -       &  -0.067 & 0.048 & 0.228 & \textbf{0.518} & \textbf{0.801} & 0.204 \\
GoTo    & -       & -       & -       & -       & -       & 0.180   & -0.066       \\
発熱 (Fever)    & 0.031   & 0.101   & -0.438  & -       & -       & -       & 0.266       \\
死者 (Fatality) & \textbf{0.327}   & 0.027   & \textbf{0.328}   & -0.080  & 0.210   & -0.455  & -0.168    \\
ALL     & 0.168   & 0.292   & 0.148   & 0.041   & 0.189   & 0.150   & 0.298    \\ \hline
\end{tabular}
\end{table*}
\end{CJK}

First, the score for ``ALL'' is below 0.3 for the entire period, although there are some fluctuations over time.
This means that no major polarization was detected for the topic of COVID-19 as a whole.
Then, for the subtopic of ``Olympic'', the controversy score is high in March and decreases in April. 
This result is interesting because the postponement of the Tokyo Olympics 2020 to 2021 was officially decided on March 24\footnote{\url{https://olympics.com/en/news/tokyo-olympic-games-postponed-ioc}}.
It is agreed that the public debate was intense in March before the official announcement was made and the controversy settled after the announcement in April.
Following this, the debate on the Olympics gained momentum again in July because the Tokyo Olympics marked one year to the 2021 Games without little sign of COVID-19 abating\footnote{\url{https://www.japantimes.co.jp/news/2020/07/10/national/politics-diplomacy/high-risk-tokyo-olympics-coronavirus/}}, which also appears to align with the higher controversy score.
We can see a gradual increase in controversy scores on vaccines from May to July.
In May and June, the government started securing the vaccines for COVID-19\footnote{\url{https://asia.nikkei.com/Business/Pharmaceuticals/Japan-in-talks-to-secure-coronavirus-vaccine-from-AstraZeneca}}, and the public debate peaked in July when the pharmaceutical companies such as Pfizer and AstraZeneca began announcing the results of their vaccine development and this was widely reported in Japan\footnote{\url{https://asia.nikkei.com/Spotlight/Coronavirus/AstraZeneca-virus-vaccine-shows-promise-in-trial}}.
Interestingly, the controversy score on vaccines only increased when people began realizing that this issue was serious, although there had been a certain volume of tweets since March.
As for Go To Travel, the number of users was smaller than expected and even the scores were less than expected.
That it became measurable in July rather than May shows that many people tweeted about their use of Go To Travel, not their opinion on it.
Also, the low scores are consistent because many people availed themselves of this economic policy and the Japan Tourism Agency reported the number of nights covered by Go To Travel was up to 88 million in five months from July 22 to 28 December\footnote{\url{https://www.mlit.go.jp/kankocho/news06_000499.html}}. This indicates that the numbers opposing this policy were minimal on Twitter.
Fever's score rises a little in August, which can be attributed to the second wave of infections in Japan\footnote{\url{https://www.nippon.com/en/japan-data/h00799/}}.
Fatality's high scores in February and April seem due to the first death by COVID-19 in Japan in February\footnote{\url{https://www.japantimes.co.jp/news/2020/02/13/national/science-health/tokyo-taxi-driver-coronavirus/}} and the government's report in April, which estimated 400,000 deaths if no action was taken\footnote{\url{https://asia.nikkei.com/Spotlight/Coronavirus/Japan-to-suffer-400-000-coronavirus-deaths-if-no-action-estimate}}.
Overall, these results confirm that the degree of controversy was aligning with the discussion in the mainstream media.

\begin{figure}[tbp]
\centering
  \subfloat[Vaccine in Jul.\label{2a}]{
      \includegraphics[width=0.28\linewidth]{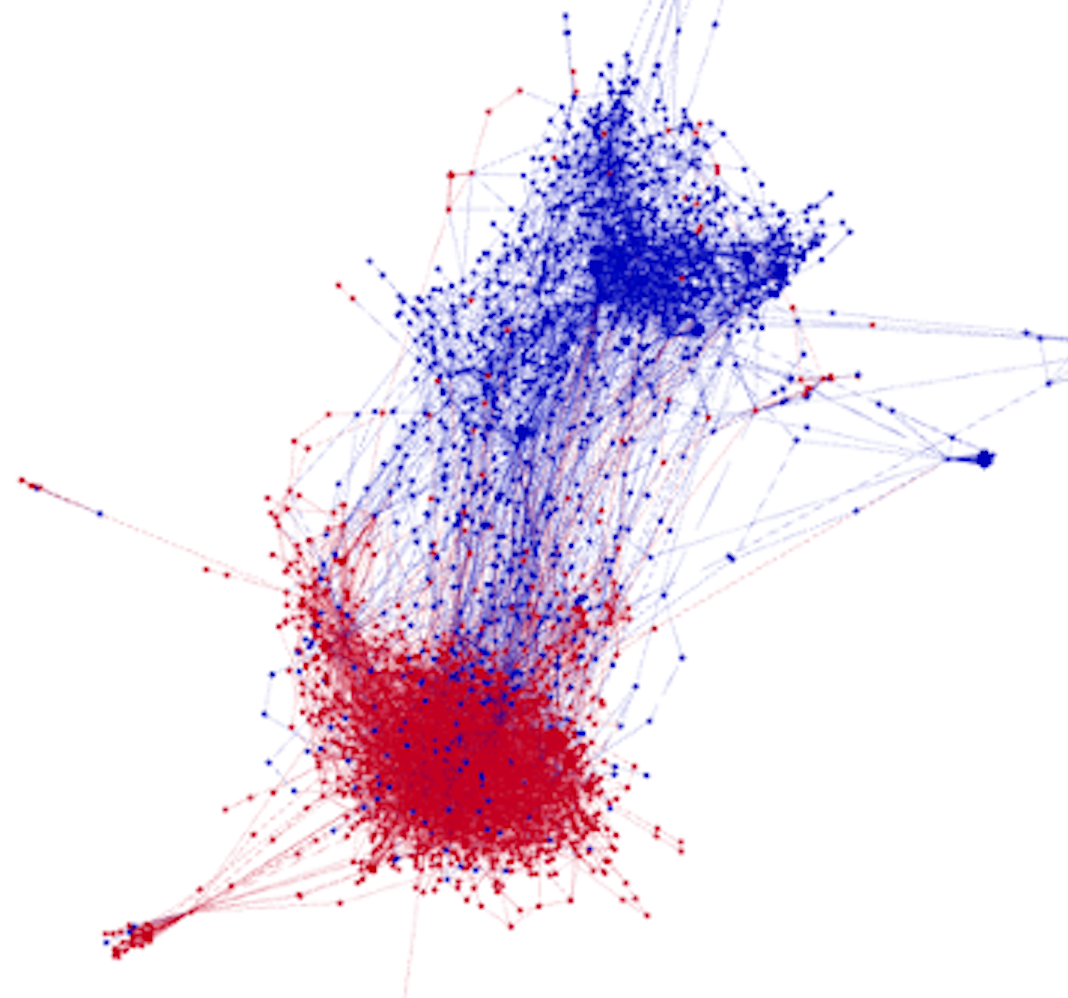}
      }
  \subfloat[Olympic in Mar.\label{2b}]{
        \includegraphics[width=0.28\linewidth]{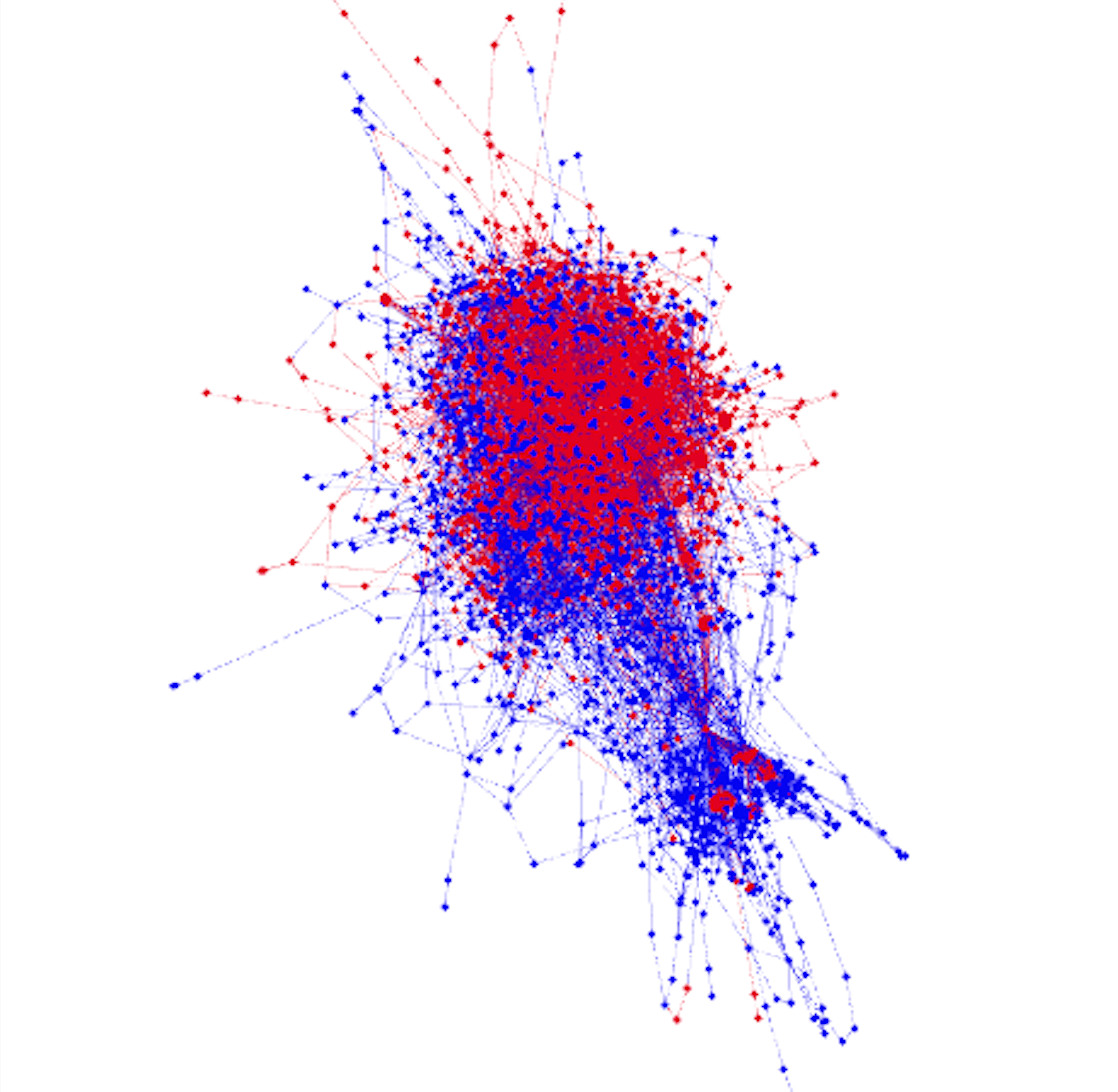}
        }
  \subfloat[Fatality in Feb.\label{2c}]{
      \includegraphics[width=0.28\linewidth]{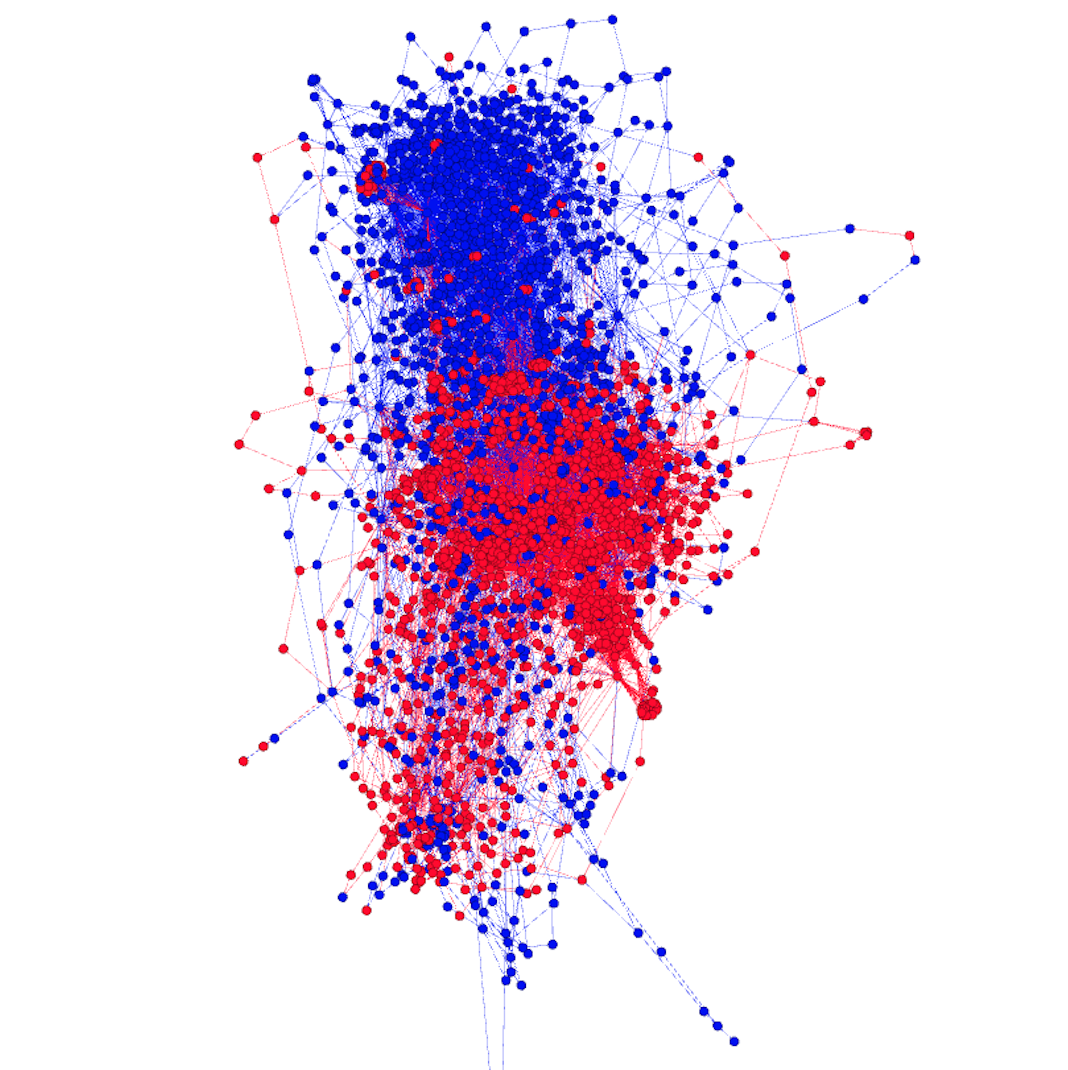}
      }
    \\
    \subfloat[Olympic in Apr.\label{2d}]{
        \includegraphics[width=0.28\linewidth]{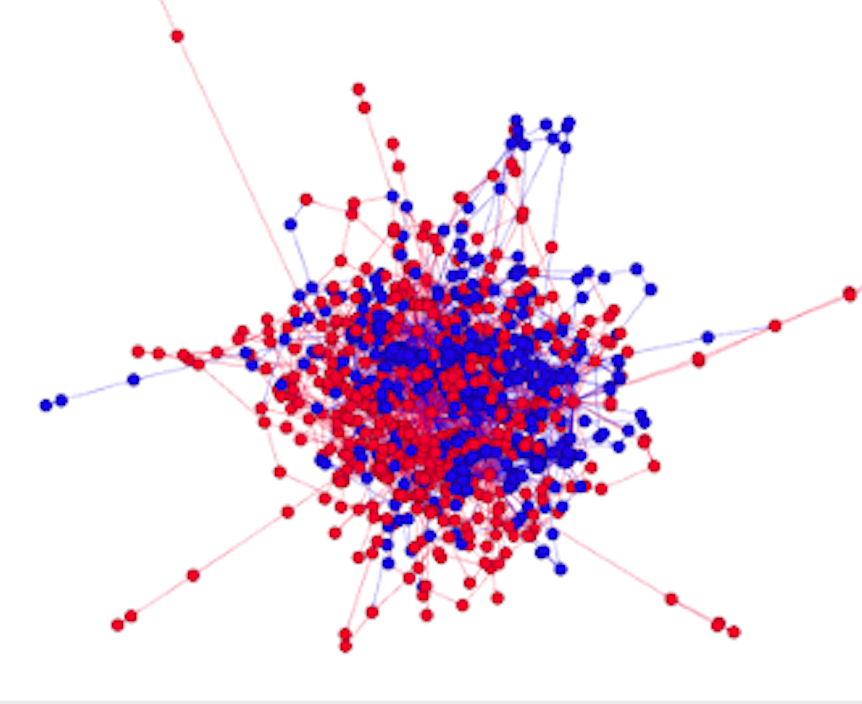}
    }
    \subfloat[GoTo in Jul.\label{2e}]{
        \includegraphics[width=0.28\linewidth]{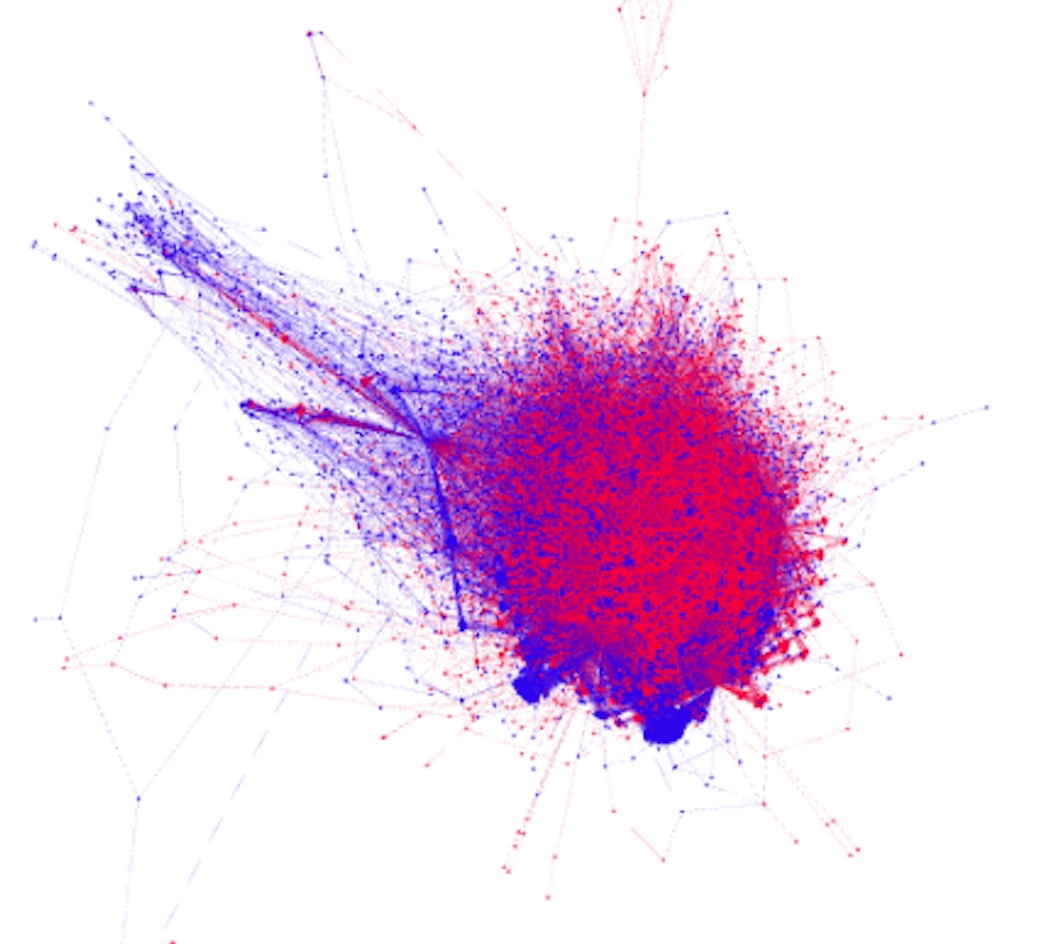}
    }
   \subfloat[Fever in Mar.\label{2f}]{
        \includegraphics[width=0.28\linewidth]{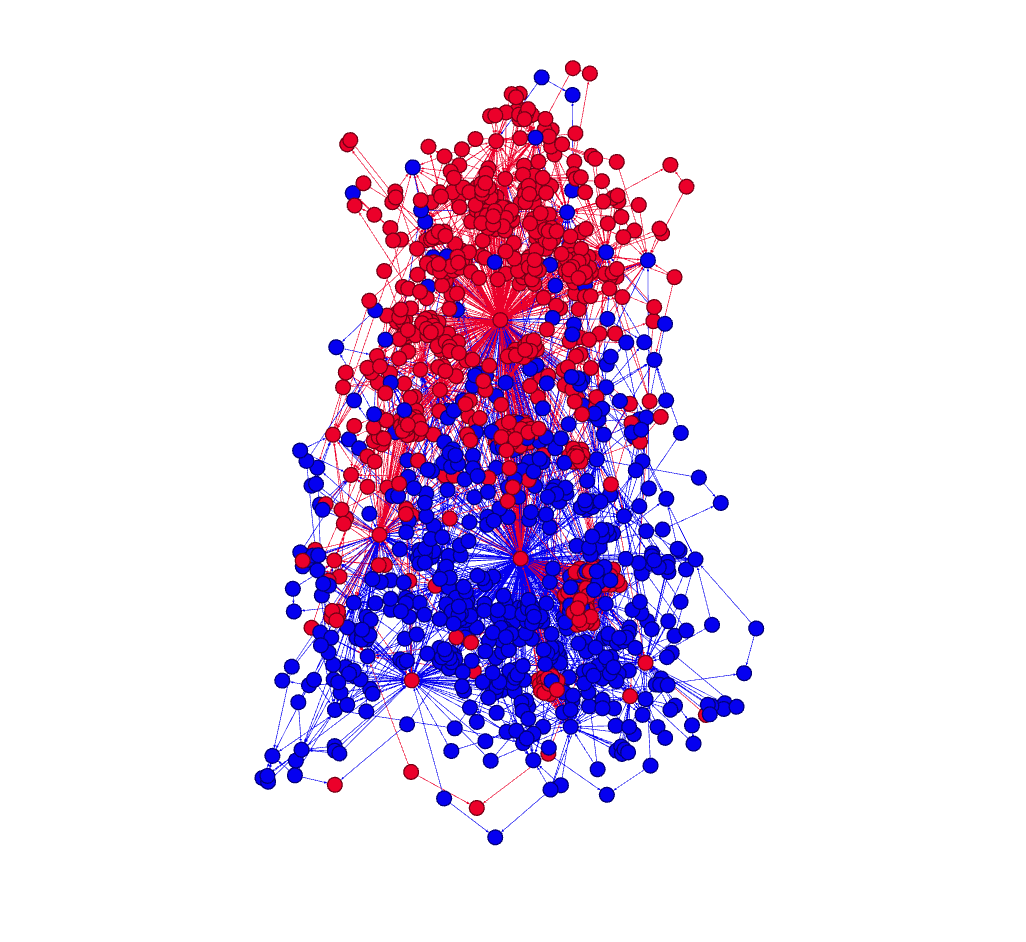}
        }

  \caption{Examples of the networks used for the calculating the controversy scores.}
\label{networks} 
\end{figure}

Additionally, in Figure~\ref{networks}, we show six examples of the networks used for calculating the controversy scores. 
The networks with high scores (subfigures~\ref{2a}~\ref{2b}~\ref{2c}) and with low scores (subfigures~\ref{2d}~\ref{2e}~\ref{2f}) were selected from Table~\ref{validation_result}.
We used the visualization tool Gephi~\cite{bastian2009gephi} with the graph layout algorithm ForceAtlas2~\cite{jacomy2014forceatlas2}.
The colors indicate the clusters divided by the METIS algorithm.
We can confirm that the two clusters have relatively small connections in the networks with a high controversy score.

\section{RQ2: Which COVID-19-related subtopics were controversial?}\label{RQ2}


\begin{CJK}{UTF8}{ipxm}
\begin{table*}[htbp]
\caption{The top 10 controversial subtopics by month with colors according to categories.}
\label{top10} 
\centering
\begin{tabular}{llllllll}
\hline
   & \multicolumn{1}{c}{Feb.}                                                      & \multicolumn{1}{c}{Mar.}                                                    & \multicolumn{1}{c}{Apr.}                                                & \multicolumn{1}{c}{May.}                                                   & \multicolumn{1}{c}{Jun.}                                               & \multicolumn{1}{c}{Jul.}                                                   & \multicolumn{1}{c}{Aug.}                                                          \\ \hline
1  & \cellcolor[HTML]{c9ace6}\begin{tabular}[c]{@{}l@{}}野党\\      (\textbf{Opposition party})\end{tabular} & 
\cellcolor[HTML]{c9ace6}\begin{tabular}[c]{@{}l@{}}首相\\      (\textbf{Prime minister})\end{tabular} &
\cellcolor[HTML]{d8f255}\begin{tabular}[c]{@{}l@{}}対象\\      (\textbf{Subjects})\end{tabular}   &
\cellcolor[HTML]{bfe4ff}\begin{tabular}[c]{@{}l@{}}学校\\      (\textbf{School})\end{tabular}        &
\cellcolor[HTML]{bfe4ff}\begin{tabular}[c]{@{}l@{}}学校\\      (\textbf{School})\end{tabular}    &
\cellcolor[HTML]{ffcabf}\begin{tabular}[c]{@{}l@{}}ワクチン\\      (\textbf{Vaccine})\end{tabular}     &
\cellcolor[HTML]{c9ace6}\begin{tabular}[c]{@{}l@{}}知事\\      (\textbf{Governor})\end{tabular}            \\
2  & \cellcolor[HTML]{c9ace6}\begin{tabular}[c]{@{}l@{}}方針\\      (\textbf{Policy})\end{tabular}           &
\cellcolor[HTML]{d8f255}\begin{tabular}[c]{@{}l@{}}ライブ\\      (\textbf{Live})\end{tabular}          &
\cellcolor[HTML]{ffcabf}\begin{tabular}[c]{@{}l@{}}医療\\      (\textbf{Medical})\end{tabular}    &
\cellcolor[HTML]{c9ace6}\begin{tabular}[c]{@{}l@{}}議員\\      (\textbf{Diet member})\end{tabular} &
\cellcolor[HTML]{c9ace6}\begin{tabular}[c]{@{}l@{}}知事\\      (\textbf{Governor})\end{tabular}  &
\cellcolor[HTML]{c9ace6}\begin{tabular}[c]{@{}l@{}}政府\\      (\textbf{Government})\end{tabular}    &
\cellcolor[HTML]{ffcabf}\begin{tabular}[c]{@{}l@{}}専門家\\      (\textbf{Expert})\end{tabular}              \\
3  & \cellcolor[HTML]{c9ace6}\begin{tabular}[c]{@{}l@{}}桜\\      (\textbf{Cherry blossom})\end{tabular}    &
\cellcolor[HTML]{c9ace6}\begin{tabular}[c]{@{}l@{}}政府\\      (\textbf{Government})\end{tabular}     &
\cellcolor[HTML]{c9ace6}\begin{tabular}[c]{@{}l@{}}安倍\\      (\textbf{Abe})\end{tabular}        &
\cellcolor[HTML]{bfe4ff}\begin{tabular}[c]{@{}l@{}}先生\\      (\textbf{Teacher})\end{tabular}       &
\cellcolor[HTML]{ffcabf}\begin{tabular}[c]{@{}l@{}}ワクチン\\      (\textbf{Vaccine})\end{tabular} &
\cellcolor[HTML]{ffcabf}\begin{tabular}[c]{@{}l@{}}陰性\\      (\textbf{Negative case})\end{tabular} &
\cellcolor[HTML]{c9ace6}\begin{tabular}[c]{@{}l@{}}安倍首相\\      (\textbf{PM Abe})\end{tabular} \\
4  &
\cellcolor[HTML]{c9ace6}\begin{tabular}[c]{@{}l@{}}政府\\      (\textbf{Government})\end{tabular}       & 
\cellcolor[HTML]{c9ace6}\begin{tabular}[c]{@{}l@{}}方針\\      (\textbf{Policy})\end{tabular}         &
\cellcolor[HTML]{ffcabf}\begin{tabular}[c]{@{}l@{}}心\\      (\textbf{Mental})\end{tabular}        &
\cellcolor[HTML]{bfe4ff}\begin{tabular}[c]{@{}l@{}}オンライン\\      (\textbf{Online})\end{tabular}     &
\cellcolor[HTML]{ffcabf}\begin{tabular}[c]{@{}l@{}}風邪\\      (\textbf{Cold})\end{tabular}      &
\cellcolor[HTML]{ffcabf}\begin{tabular}[c]{@{}l@{}}重症\\      (\textbf{Severe case})\end{tabular}   &
\cellcolor[HTML]{d8f255}\begin{tabular}[c]{@{}l@{}}企業\\      (\textbf{Company})\end{tabular}              \\
5  & 
\cellcolor[HTML]{ffcabf}\begin{tabular}[c]{@{}l@{}}医療\\      (\textbf{Medical})\end{tabular}                   &
\cellcolor[HTML]{d8f255}\begin{tabular}[c]{@{}l@{}}経済\\      (\textbf{Economy})\end{tabular}        &
\cellcolor[HTML]{c9ace6}\begin{tabular}[c]{@{}l@{}}政府\\      (\textbf{Government})\end{tabular} &
\cellcolor[HTML]{ffcabf}\begin{tabular}[c]{@{}l@{}}専門家\\      (\textbf{Expert})\end{tabular}       &
\cellcolor[HTML]{c8c8cb}\begin{tabular}[c]{@{}l@{}}社会\\      (\textbf{Society})\end{tabular}   &
\cellcolor[HTML]{bfe4ff}\cellcolor[HTML]{bfe4ff}\begin{tabular}[c]{@{}l@{}}学校\\      (\textbf{School})\end{tabular}        &
\cellcolor[HTML]{d8f255}\cellcolor[HTML]{d8f255}\begin{tabular}[c]{@{}l@{}}経済\\      (\textbf{Economy})\end{tabular}              \\
6  &
\cellcolor[HTML]{c8c8cb}\begin{tabular}[c]{@{}l@{}}死者\\      (\textbf{Fatality})\end{tabular}                  &
\cellcolor[HTML]{ffcabf}\begin{tabular}[c]{@{}l@{}}重症\\      (\textbf{Severe case})\end{tabular}    & 
\cellcolor[HTML]{c8c8cb}\begin{tabular}[c]{@{}l@{}}死者\\      (\textbf{Fatality})\end{tabular}               &
\cellcolor[HTML]{c8c8cb}\begin{tabular}[c]{@{}l@{}}市民\\      (\textbf{Citizen})\end{tabular}       &
\begin{tabular}[c]{@{}l@{}}都民\\      (Tokyo citizen)\end{tabular}  & 
\cellcolor[HTML]{d8f255}\begin{tabular}[c]{@{}l@{}}オリンピック\\      (\textbf{Olympic})\end{tabular}      &
\cellcolor[HTML]{ffcabf}\begin{tabular}[c]{@{}l@{}}風邪\\      (\textbf{Cold})\end{tabular}                 \\
7  &
\begin{tabular}[c]{@{}l@{}}予算\\      (Budget)\end{tabular}                    &
\cellcolor[HTML]{d8f255}\begin{tabular}[c]{@{}l@{}}オリンピック\\      (\textbf{Olympic})\end{tabular}             &
\begin{tabular}[c]{@{}l@{}}無料\\      (Free)\end{tabular}                &
\begin{tabular}[c]{@{}l@{}}首相\\      (Prime minister)\end{tabular}         & \begin{tabular}[c]{@{}l@{}}安倍政権\\      (Abe admin)\end{tabular}                                                                   & 
\cellcolor[HTML]{c8c8cb}\begin{tabular}[c]{@{}l@{}}災害\\      (\textbf{Disaster})\end{tabular}            &
\begin{tabular}[c]{@{}l@{}}熱中症\\      (Heatstroke)\end{tabular}                   \\
8  & 
\begin{tabular}[c]{@{}l@{}}ネット\\      (Internet)\end{tabular}                 &
\cellcolor[HTML]{d8f255}\begin{tabular}[c]{@{}l@{}}消費税\\      (\textbf{Consumption Tax})\end{tabular}        &
\begin{tabular}[c]{@{}l@{}}会社\\      (Company)\end{tabular}             &
\begin{tabular}[c]{@{}l@{}}安倍政権\\      (Abe admin)\end{tabular}   & 
\begin{tabular}[c]{@{}l@{}}医師\\      (Medical Doctor)\end{tabular}     &
\cellcolor[HTML]{c8c8cb}\begin{tabular}[c]{@{}l@{}}雨\\      (\textbf{Rain})\end{tabular}    & 
\begin{tabular}[c]{@{}l@{}}医師\\      (Medical Doctor)\end{tabular}                \\
9  & \begin{tabular}[c]{@{}l@{}}インフル\\      (Influenza)\end{tabular}               &
\begin{tabular}[c]{@{}l@{}}リスク\\      (Risk)\end{tabular}                   & \begin{tabular}[c]{@{}l@{}}医師\\      (Medical Doctor)\end{tabular}      & 
\begin{tabular}[c]{@{}l@{}}危機\\      (Crisis)\end{tabular}                & \begin{tabular}[c]{@{}l@{}}命\\      (Life)\end{tabular}                &  
\begin{tabular}[c]{@{}l@{}}安倍政権\\      (Abe admin)\end{tabular} & \begin{tabular}[c]{@{}l@{}}休み\\      (Holiday)\end{tabular}                       \\
10 & \begin{tabular}[c]{@{}l@{}}インフルエンザ\\      (Influenza)\end{tabular}            &
\begin{tabular}[c]{@{}l@{}}マスク\\      (Mask)\end{tabular}                   &
\begin{tabular}[c]{@{}l@{}}外出自粛\\(Self-quarantine)\end{tabular}                                                         & \begin{tabular}[c]{@{}l@{}}ワクチン\\      (Vaccine)\end{tabular}              &
\begin{tabular}[c]{@{}l@{}}経済\\      (Economy)\end{tabular}            &
\begin{tabular}[c]{@{}l@{}}キャンペーン\\      (Campaign)\end{tabular}          &
\begin{tabular}[c]{@{}l@{}}ワクチン\\      (Vaccine)\end{tabular}                     \\ \hline

\end{tabular}
\end{table*}
\end{CJK}

As reported in the following sections, we did not pre-specify the subtopics but automatically extracted the keywords indicating subtopics.
Table~\ref{top10} shows the top 10 controversial subtopics extracted by the proposed framework for each month.
We added the English translation of each word (shown in brackets).
The subtopics with controversy scores of more than 0.3 are shown in bold.
For ones that exceeded the threshold, we also colored them according to the larger category we recognized: purple for government, pink for medical issues, green for economy, blue for education, and gray for others.

This table shows how controversies transition over time: 
governmental issues/policies are the main controversies in February (``Cherry  blossom'' indicates the cherry blossom viewing party, where immense suspicion revolved around the ex-prime minister Abe with regard to inviting antisocial elements to his party\footnote{\url{https://www.japantimes.co.jp/news/2019/11/27/reference/cherry-blossom-viewing-party-shinzo-abe-cronyism-scandal/}}); 
the concerns around economic activities emerged in March (``Live'' and ``Economy'');
medical concerns emerged in April (``Medical'' and ``Mental'') as well as economic issues (``Subjects'' indicate the subjects of the government's support policy);
the controversy on education issues exacerbated in May (``School,'' ``Teacher'' and ``Online''), and the medical experts meeting for COVID-19 received attention (``Expert'');
the news on vaccines began going viral in June as stated in the previous section (``Vaccine'');
the disastrous rain that hit southern Japan\footnote{\url{https://asia.nikkei.com/Economy/Natural-disasters/Torrential-rain-triggers-massive-flooding-in-southwestern-Japan}} in July (``Disaster'' and ``Rain'');
the controversial subtopics in August were already noticed in the previous months.

In general, governmental issues always top the list.
Also, we could see some significant subtopics associated with the terms economic issues, medical issues including vaccines, and educational issues.
These controversial subtopics all conform to the public debates in the news media.
Interestingly, each subtopic tops the list in turn.
It was considered whether the shifting of people's interests produced general patterns, although this point could not be tested.

\section{RQ3: What was the scale and sentiment of the controversial subtopics?}\label{RQ3}

\subsection{The correlation of controversy score with the scale and sentiment of subtopics}
\begin{figure*}[htbp]
\centering
  \subfloat[Number of users\label{3a}]{
      \includegraphics[width=0.28\linewidth]{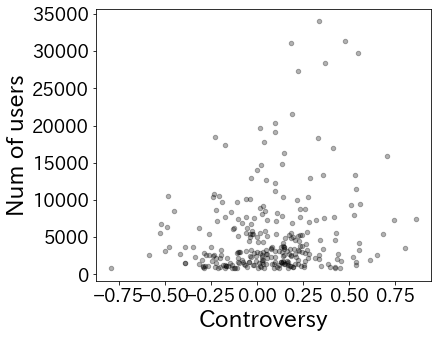}
      }
  \subfloat[Sentiment mean\label{3b}]{
        \includegraphics[width=0.28\linewidth]{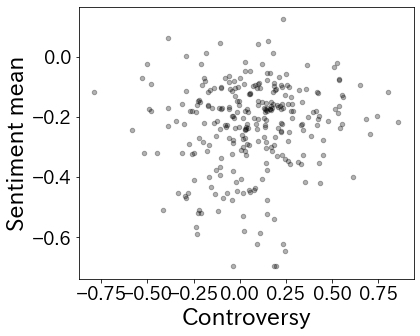}
        }
  \subfloat[Sentiment std\label{3c}]{
        \includegraphics[width=0.28\linewidth]{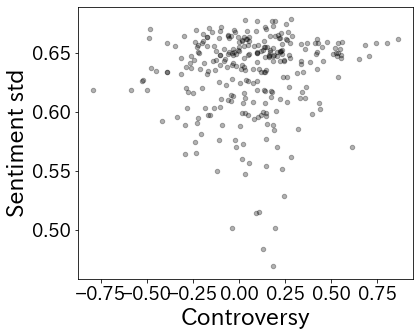}
        }
  \caption{Comparison of controversy score with the number of users and sentiments (mean and standard deviation (std)).}
\label{scatter} 
\end{figure*}

In this RQ, we analyze the subtopics with a high controversy score, using the traditional indicators of scale and sentiment, which are the de facto standard measures for social media analysis~\cite{agarwal2011sentiment}.
For this purpose, we drew scatter plots shown in Figure~\ref{scatter}, in which each dot indicates a subtopic in a month.
The number of users indicates the number of nodes for calculating the controversy scores for subtopics, which also indicates how substantial people's attention gathered to the subtopics.
We also calculated the mean and standard deviation (std) of the sentiment for each subtopic. 
Here, we used the standard deviation, not only the mean, because we assumed the sentiment would be polarized by clusters, i.e., positive on one side and negative on the other.
We calculated the sentiment scores using Oseti\footnote{\url{https://github.com/ikegami-yukino/oseti} [accessed: 2021/05]}, a Japanese sentiment analyzer using major Japanese dictionaries~\cite{kobayashi2004collecting,higashiyama2008learning}.
We note that we also tried the BERT-based sentiment analysis, but this did not affect the overall result in the paper.

We do not discern strong correlations in these scatter plots.
We calculated each correlation coefficient with the controversy scores and obtained 0.186 ($p$=0.002), 0.086 ($p$=0.013), and 0.039 ($p$=0.020) respectively for the number of users, sentiment mean, and sentiment std.
This result indicates that the controversy score is independent of these traditional indicators for social media analysis; hence this score offers another dimension for understanding the discussions taking place on social media.
Primarily, it is interesting that the sentiment std and the controversy score are not strongly correlated. 
This implies that polarization that takes the form of an echo chamber is hard to detect by using only sentiment scores, which appears to enhance the value of the controversy quantification method.

\subsection{The relationship between the scale and controversy scores for subtopics}
In the previous subsection, we found little correlation between the size and the controversy of subtopics.
This indicates that polarization does not necessarily occur in subtopics that attract the considerable attention of the public.
In this subsection, we further compare and analyze large subtopics with high controversy scores and ones with low controversy scores.
As before, the controversy threshold is set to 0.3, and we set the topic threshold size as 10,000 by referring to Figure~\ref{3a}. 
Table~\ref{size_con} shows, for each month, the subtopics with users of 10,000, divided into groups with controversy scores greater than 0.3 and those with controversy scores less than 0.3.
For visibility, we only included translated English words in the table.

\begin{table}[htbp]
\centering
\caption{Subtopics with a large number of users divided into a high controversy group and low controversy group.}
\label{size_con} 
\begin{tabular}{cccc}
\hline
                      & \multicolumn{3}{c}{Subtopics}                                                                                                                                                                                                                                                   \\ \cline{2-4} 
Month                 & High controversy                                                                           & \multicolumn{2}{c}{Low controversy}                                                                                                                                                \\ \hline \hline
Feb.                  & Government                                                                                 & Cruise ship                                                                                        & Abe                                                                           \\ \hline 
Mar.                  & \begin{tabular}[c]{@{}c@{}}Government\\      Economy\\      Prime minister\end{tabular}    & \begin{tabular}[c]{@{}c@{}}Opposition party\\      Medical\end{tabular}                            & Abe                                                                           \\ \hline
Apr.                  & \begin{tabular}[c]{@{}c@{}}Government\\      Medical\\      Abe\\      Mental\end{tabular} & \begin{tabular}[c]{@{}c@{}}Mask\\      Fatality\\      State of emergency\\      Life\end{tabular} & \begin{tabular}[c]{@{}c@{}}Economy\\      Crisis\\      Governor\end{tabular} \\ \hline
\multirow{2}{*}{May.} & \multirow{2}{*}{-}                                                                         & \begin{tabular}[c]{@{}c@{}}Government\\      Medical\\      Abe\end{tabular}                       & \begin{tabular}[c]{@{}c@{}}Economy\\ Admin\\ Abe admin\end{tabular}           \\
                      &                                                                                            & \multicolumn{2}{c}{Public Prosecutors Office law}                                                                                                                                  \\ \hline
Jun,                  & Governor                                                                                   & -                                                                                                  &                                                                               \\ \hline
Jul.                  & Government                                                                                 & \begin{tabular}[c]{@{}c@{}}Medical\\      Governor\end{tabular}                                    & \begin{tabular}[c]{@{}c@{}}Campaign\\      Record high\end{tabular}           \\ \hline
Aug.                  & -                                                                                          & \begin{tabular}[c]{@{}c@{}}Government\\      Diet\end{tabular}                                     & Abe                                                                           \\ \hline
\end{tabular}
\end{table}

As seen in the previous section, the high controversy group contains governmental subtopics and some economic and medical topics.
The government, economy, and medical subtopics can also be seen in the low controversy group, which indicates that the controversy scores depend on time.
Some subtopics are exclusively found in the low controversy group.
For example, the cruise ship subtopic in February dealt with the luxury cruise ship Diamond Princess\footnote{\url{https://edition.cnn.com/2020/02/04/asia/coronavirus-japan-cruise-intl-hnk/index.html}}, which was anchored on Japanese shores during a round-the-world trip, and COVID-19 infections were found onboard.
This became a major topic of discussion on whether to disembark the infected people because infection control measures were lacking in Japan during the time.
Nevertheless, it did not become a controversial topic (controversy score=-0.208) because few opinions on social media strongly promoted disembarking the passengers.
Next, ``Mask'' in April means ``Abenomask'', a policy announced by ex-prime minister Abe, which was to distribute two masks to all households, but it was criticized because of the vast amount of money spent for it\footnote{\url{https://asia.nikkei.com/Spotlight/Coronavirus/Will-Abenomasks-help-prevent-coronavirus-spread-in-Japan}}.
Abenomask did not become highly controversial because there was no strong support for it on social media (controversy score=0.189).
``Public Prosecutors Office law'' was intensively discussed in May, but its controversy score was low (0.022). 
Ex-prime minister Abe sought to extend the term of office of the superintending prosecutor of the Tokyo High Public Prosecutors Office, which was regarded as a forced measure to advance Abe's political agenda significantly. 
In addition, it was discovered that the prosecutor was gambling at mahjong with news writers, without social distancing, during the COVID-19 pandemic, which led to his resignation and further controversy\footnote{\url{https://mainichi.jp/english/articles/20200522/p2a/00m/0na/012000c}}, which fueled the discussion.
While ``Cherry blossom'' described in the previous section was also a scandal of sorts and the associated controversy extensive, the prosecutor's case was less controversial.
The difference is that "Cherry blossom" was only a suspicion, while the prosecutor's case overwhelmingly aroused public opposition, and the polarization was low.
The ``Campaign'' in July indicates ``Go To Travel'', and as already mentioned, the controversy score was not significantly high for this subtopic (controversy score=0.219).

This result shows that public criticism does not necessarily increase the controversy score.
For example, while Mask and Prosecutors received extensive criticism, their controversy scores were not high.
This appears so because the criticism was overwhelming, and there were few positive opinions on these subtopics thus polarization did not occur.
In a sense, the controversy score is not suitable for the government for extracting only critical opinions from social media, not polarization.

\subsection{The relationship between the sentiment and controversy scores for subtopics}
Next, we look at the relationship between sentiment and controversy for the subtopics by investigating the controversy scores for the subtopics with low sentiment.
However, Figure~\ref{3b} shows that when the level of low sentiment is less than -0.5 (the threshold we set by referring to Figure~\ref{3b}), there are no subtopics with a high controversy score.
Therefore, we examine all the subtopics with sentiment below -0.5 for each month in Table~\ref{senti_con}.

\begin{table}[htbp]
\centering
\caption{Subtopics with a low sentiment score (all have a low controversy score).}
\label{senti_con} 
\begin{tabular}{lc}
\hline
Month & Subtopcics                                                                                                    \\ \hline\hline
Feb.  & \begin{tabular}[c]{@{}c@{}}Fatality\\      Cruise ship\\      Passenger\end{tabular}                          \\ \hline
Mar.  & Fatality                                                                                                      \\ \hline
Apr.  & Fatality                                                                                                      \\ \hline
May.  & Fatality                                                                                                      \\ \hline
Jun.  & Fatality                                                                                                      \\ \hline
Jul.  & \begin{tabular}[c]{@{}c@{}}Fatality\\      State of emergency\\      Record high\\      Employee\end{tabular} \\ \hline
Aug.  & Fatality                                                                                                      \\ \hline
\end{tabular}
\end{table}

``Fatality'' is observed to be present every month.
This implies that strong negative sentiments always accompany reports of fatalities and the comments on them.
In a sense, this result demonstrates one of the limitations of sentiment analysis; it picks up any negative words without deep consideration of context, which may not aid political decision-making.
Next, the topic of ``Cruise ship'' and ``Passenger'' are included in February, which are indicators of the Diamond Princess issue being discussed.
This was the first major COVID-19 related threat for Japan, and many negative comments were issued, indicating that sentiment was low.
Also, a topic regarding the declaration of a state of emergency was included in July.
This indicates that with the number of COVID-19 cases on the rise, there was widespread speculation of a second emergency declaration and a strong disapproval for the same.

This result confirms that the low sentiment score identifies subtopics with a negative impact on people.
However, we confirmed that factual words such as ``Fatality'' were also extracted with low sentiment scores in the context of the pandemic contained little relevant information.

\section{Discussion}
\subsection{Summary of controversial subtopics on COVID-19 in Japan}
We explored the subtopics of COVID-19 on controversy according to the proposed framework without prior knowledge, and the individual subtopics were shown in Section~\ref{RQ2} and~\ref{RQ3}.
Although not statistically processed, the results indicate a tendency for the main controversial subtopics to be the government, medical matters, the economy, and education.
Conversely, we also obtained subtopics that have received more public attention but have not become more controversial.
These differences can be attributed to the following factors.

\paragraph{Criticism of a specific person}
Government criticism, particularly the subtopics about prime ministers and governors is central to major controversies.
Conversely, subtopics concerning policy did not appear to be subject to major controversies.
For example, policies such as the distribution of masks to all households and Go To Travel had a low controversy score. Simultaneously, the controversy score for politicians was high because of their fame and for their associated scandals (e.g., cherry blossom).
This may be because the public tends to discuss politicians’ morality rather than their policies.
As there may be different responses between countries, it is desirable to investigate subtopics in different countries. 

\paragraph{Familiar subtopics}
Medical matters, economics, and education are particularly appealing to the public.
In particular, we found subtopics on vaccines, mental health, restaurant closures, and children's school attendance to be major controversies.
It is easy for people to express their opinions on such subtopics.
Against this, subtopics such as cruise ships were less controversial.
This may be because people felt such topics were local occurrences and unfamiliar issues.

\paragraph{Topics that have not yet been concluded}
The inconclusive issues are likely to get high controversy scores such as Cherry blossoms or the Olympic game before the announcement of its postponement.
Conversely, prosecutor scandal was not polarized because the opinion of one side was overwhelming. 
Also, the polarization of the Olympics was settled after the announcement.

\subsection{Implications for decision makers}
We compared controversy with sentiment and topic scale and found that controversy can be an indicator of an aspect that does not correlate with either.
We hope that the use of this controversy score will be discussed more deeply in the context of evidence-based policy-making (EBPM)~\cite{behague2009evidence} or marketing~\cite{woerdl2008internet} in the future.
In addition, although we did not conduct a deeper analysis of what is discussed in the polarized subtopic, future researchers or practitioners can analyze the contents of subtopics using methods proposed in previous research~\cite{jang2018explaining} such as network extractive summarization which aims to find tweets that are representative of clusters.

\subsection{Versatility of this framework}
We used Twitter in this study.
Twitter is indeed an appropriate platform for quantifying echo chambers because it is easy to create endorsement networks there.
However network-based echo chambers can be created on other platforms (e.g., Facebook~\cite{johnson2020online}), which enables the proposed framework to be used there.
Also, as the previous research mentioned~\cite{garimella2018quantifying}, this method is independent of languages.

\subsection{Limitations of the proposed framework}
One of the major limitations of this framework is that it requires a certain number of users to measure the controversy score.
Therefore, it is difficult to apply this method to topics that are too small.
However, with sufficient data, analysis can be conducted on a smaller time scale instead of the one-month window that we set to compare subtopics.
Also, since we measure the distance between clusters in the network, two clusters may not necessarily disagree.
For example, when two clusters occur, both can be negative towards the mainstream but negative for different reasons such as being critical of a policy instead of a political personality.

\section{Conclusion}
In this study, we proposed a framework for discovering controversial subtopics in social media.
In the experiment, we specified well-known subtopics related to COVID-19 in Japan and examined these controversy scores and their transitions over time.
The results confirmed that the framework captured the controversial subtopics that well reflect current reality.
In subsequent analysis, we used the framework without pre-specified subtopics to examine subtopics with high controversy scores.
The results showed that subtopics on government, medical matters, economy, and education had high controversy scores.
We also compared the scores with the scale and sentiment of the subtopics and found that the controversy score is minimally correlated with these traditional indicators. If we assume that public polarization is an important political topic, the controversy score can be one crucial indicator for gauging its influence.
This study broadens the horizon of existing social media analysis and provides deeper insights for future research into social media.



\bibliographystyle{IEEEtran}
\bibliography{IEEEabrv,IEEEbib}

\begin{thebibliography}{10}
\providecommand{\url}[1]{#1}
\csname url@samestyle\endcsname
\providecommand{\newblock}{\relax}
\providecommand{\bibinfo}[2]{#2}
\providecommand{\BIBentrySTDinterwordspacing}{\spaceskip=0pt\relax}
\providecommand{\BIBentryALTinterwordstretchfactor}{4}
\providecommand{\BIBentryALTinterwordspacing}{\spaceskip=\fontdimen2\font plus
\BIBentryALTinterwordstretchfactor\fontdimen3\font minus
  \fontdimen4\font\relax}
\providecommand{\BIBforeignlanguage}[2]{{%
\expandafter\ifx\csname l@#1\endcsname\relax
\typeout{** WARNING: IEEEtran.bst: No hyphenation pattern has been}%
\typeout{** loaded for the language `#1'. Using the pattern for}%
\typeout{** the default language instead.}%
\else
\language=\csname l@#1\endcsname
\fi
#2}}
\providecommand{\BIBdecl}{\relax}
\BIBdecl

\bibitem{burki2019vaccine}
T.~Burki, ``Vaccine misinformation and social media,'' \emph{The Lancet Digital
  Health}, vol.~1, no.~6, pp. e258--e259, 2019.

\bibitem{national1989improving}
N.~R. Council \emph{et~al.}, ``Improving risk communication,'' 1989.

\bibitem{conover2011political}
M.~Conover, J.~Ratkiewicz, M.~Francisco, B.~Gon{\c{c}}alves, F.~Menczer, and
  A.~Flammini, ``Political polarization on twitter,'' in \emph{Proceedings of
  the International AAAI Conference on Web and Social Media}, vol.~5, no.~1,
  2011.

\bibitem{schmidt2018polarization}
A.~L. Schmidt, F.~Zollo, A.~Scala, C.~Betsch, and W.~Quattrociocchi,
  ``Polarization of the vaccination debate on facebook,'' \emph{Vaccine},
  vol.~36, no.~25, pp. 3606--3612, 2018.

\bibitem{williams2015network}
H.~T. Williams, J.~R. McMurray, T.~Kurz, and F.~H. Lambert, ``Network analysis
  reveals open forums and echo chambers in social media discussions of climate
  change,'' \emph{Global environmental change}, vol.~32, pp. 126--138, 2015.

\bibitem{makridis2020real}
C.~Makridis and J.~T. Rothwell, ``The real cost of political polarization:
  evidence from the covid-19 pandemic,'' \emph{Available at SSRN 3638373},
  2020.

\bibitem{sasahara2021social}
K.~Sasahara, W.~Chen, H.~Peng, G.~L. Ciampaglia, A.~Flammini, and F.~Menczer,
  ``Social influence and unfollowing accelerate the emergence of echo
  chambers,'' \emph{Journal of Computational Social Science}, vol.~4, no.~1,
  pp. 381--402, 2021.

\bibitem{nelimarkka2019re}
M.~Nelimarkka, J.~P. Rancy, J.~Grygiel, and B.~Semaan, ``(re) design to
  mitigate political polarization: Reflecting habermas' ideal communication
  space in the united states of america and finland,'' \emph{Proceedings of the
  ACM on Human-Computer Interaction}, vol.~3, no. CSCW, pp. 1--25, 2019.

\bibitem{garimella2018quantifying}
K.~Garimella, G.~D.~F. Morales, A.~Gionis, and M.~Mathioudakis, ``Quantifying
  controversy on social media,'' \emph{ACM Transactions on Social Computing},
  vol.~1, no.~1, pp. 1--27, 2018.

\bibitem{guerra2013measure}
P.~Guerra, W.~Meira~Jr, C.~Cardie, and R.~Kleinberg, ``A measure of
  polarization on social media networks based on community boundaries,'' in
  \emph{Proceedings of the International AAAI Conference on Web and Social
  Media}, vol.~7, no.~1, 2013.

\bibitem{gillani2018me}
N.~Gillani, A.~Yuan, M.~Saveski, S.~Vosoughi, and D.~Roy, ``Me, my echo
  chamber, and i: introspection on social media polarization,'' in
  \emph{Proceedings of the 2018 World Wide Web Conference}, 2018, pp. 823--831.

\bibitem{coletto2017motif}
M.~Coletto, K.~Garimella, A.~Gionis, and C.~Lucchese, ``A motif-based approach
  for identifying controversy,'' in \emph{Proceedings of the International AAAI
  Conference on Web and Social Media}, vol.~11, no.~1, 2017.

\bibitem{woerdl2008internet}
M.~Woerdl, S.~Papagiannidis, M.~A. Bourlakis, and F.~Li, ``Internet-induced
  marketing techniques: Critical factors in viral marketing campaigns.''
  \emph{Journal of Business Science and Applied Management}, vol.~3, no.~1, pp.
  35--45, 2008.

\bibitem{batrinca2015social}
B.~Batrinca and P.~C. Treleaven, ``Social media analytics: a survey of
  techniques, tools and platforms,'' \emph{Ai \& Society}, vol.~30, no.~1, pp.
  89--116, 2015.

\bibitem{murdock2015visualization}
J.~Murdock and C.~Allen, ``Visualization techniques for topic model checking,''
  in \emph{Proceedings of the AAAI Conference on Artificial Intelligence},
  vol.~29, no.~1, 2015.

\bibitem{kudo2006mecab}
T.~Kudo, ``Mecab: Yet another part-of-speech and morphological analyzer,''
  \emph{http://mecab. sourceforge. jp}, 2006.

\bibitem{sato2017implementation}
T.~Sato, T.~Hashimoto, and M.~Okumura, ``Implementation of a word segmentation
  dictionary called mecab-ipadic-neologd and study on how to use it effectively
  for information retrieval,'' in \emph{Proceedings of the twenty-three annual
  meeting of the association for natural language processing}.\hskip 1em plus
  0.5em minus 0.4em\relax The Association for Natural Language Processing,
  2017, pp. NLP2017--B6.

\bibitem{hong2010empirical}
L.~Hong and B.~D. Davison, ``Empirical study of topic modeling in twitter,'' in
  \emph{Proceedings of the first workshop on social media analytics}, 2010, pp.
  80--88.

\bibitem{karypis1997metis}
G.~Karypis and V.~Kumar, ``Metis: A software package for partitioning
  unstructured graphs, partitioning meshes, and computing fill-reducing
  orderings of sparse matrices,'' 1997.

\bibitem{alvarez2006large}
J.~I. Alvarez-Hamelin, L.~Dall'Asta, A.~Barrat, and A.~Vespignani, ``Large
  scale networks fingerprinting and visualization using the k-core
  decomposition,'' in \emph{Advances in neural information processing systems},
  2006, pp. 41--50.

\bibitem{wec2020us}
\BIBentryALTinterwordspacing
WorldEconomicForum, ``U.s. and u.k. are optimistic indicators for covid-19
  vaccination uptake,'' December 2020, [Accecced 29-May-2021]. [Online].
  Available:
  \url{https://www.ipsos.com/en/global-attitudes-covid-19-vaccine-december-2020}
\BIBentrySTDinterwordspacing

\bibitem{bastian2009gephi}
M.~Bastian, S.~Heymann, and M.~Jacomy, ``Gephi: an open source software for
  exploring and manipulating networks,'' in \emph{Proceedings of the
  International AAAI Conference on Web and Social Media}, vol.~3, no.~1, 2009.

\bibitem{jacomy2014forceatlas2}
M.~Jacomy, T.~Venturini, S.~Heymann, and M.~Bastian, ``Forceatlas2, a
  continuous graph layout algorithm for handy network visualization designed
  for the gephi software,'' \emph{PloS ONE}, vol.~9, no.~6, p. e98679, 2014.

\bibitem{agarwal2011sentiment}
A.~Agarwal, B.~Xie, I.~Vovsha, O.~Rambow, and R.~J. Passonneau, ``Sentiment
  analysis of twitter data,'' in \emph{Proceedings of the workshop on language
  in social media (LSM 2011)}, 2011, pp. 30--38.

\bibitem{kobayashi2004collecting}
N.~Kobayashi, K.~Inui, Y.~Matsumoto, K.~Tateishi, and T.~Fukushima,
  ``Collecting evaluative expressions for opinion extraction,'' in
  \emph{International Conference on Natural Language Processing}.\hskip 1em
  plus 0.5em minus 0.4em\relax Springer, 2004, pp. 596--605.

\bibitem{higashiyama2008learning}
M.~Higashiyama, K.~Inui, and Y.~Matsumoto, ``Learning sentiment of nouns from
  selectional preferences of verbs and adjectives,'' in \emph{Proceedings of
  the 14th Annual Meeting of the Association for Natural Language Processing},
  2008, pp. 584--587.

\bibitem{behague2009evidence}
D.~Behague, C.~Tawiah, M.~Rosato, T.~Some, and J.~Morrison, ``Evidence-based
  policy-making: the implications of globally-applicable research for
  context-specific problem-solving in developing countries,'' \emph{Social
  Science \& Medicine}, vol.~69, no.~10, pp. 1539--1546, 2009.

\bibitem{jang2018explaining}
M.~Jang and J.~Allan, ``Explaining controversy on social media via stance
  summarization,'' in \emph{The 41st International ACM SIGIR Conference on
  Research \& Development in Information Retrieval}, 2018, pp. 1221--1224.

\bibitem{johnson2020online}
N.~F. Johnson, N.~Vel{\'a}squez, N.~J. Restrepo, R.~Leahy, N.~Gabriel,
  S.~El~Oud, M.~Zheng, P.~Manrique, S.~Wuchty, and Y.~Lupu, ``The online
  competition between pro-and anti-vaccination views,'' \emph{Nature}, vol.
  582, no. 7811, pp. 230--233, 2020.

\end{thebibliography}

\end{document}